\documentclass{jaa}
\usepackage{natbib}
\usepackage{graphicx}

\begin{document}\sloppy

\title{Ultraviolet spectra of comets: Rejecting the detection of pentacene, toluene, and Fe$^+$}


\author{G. ROUILL{\'E}\textsuperscript{1,*}}
\affilOne{\textsuperscript{1}Astrophysical Institute and University Observatory, Friedrich Schiller University Jena, Schillerg{\"a}sschen 2-3, D-07745 Jena, Germany.}


\twocolumn[{

\maketitle

\corres{gael.rouille@uni-jena.de}

\msinfo{1 January 2015}{1 January 2015}

\begin{abstract}
A recent study announced the detection of three bands in the ultraviolet emission spectra of more than a dozen comets, assigning two of them to pentacene (C$_{22}$H$_{14}$) and the third one to toluene (C$_7$H$_8$). The comparison of the spectra with the results of exploitable laboratory measurements on rare-gas-matrix-isolated pentacene and on jet-cooled toluene does not reveal elements that would justify the assignment, which is therefore unsubstantiated. The study also claimed the detection of an Fe~{\small{II}} line in the gas of all but one comet. Yet, spectroscopic data on Fe~{\small{II}} do not corroborate the attribution. Because spectroscopic measurements on the ultraviolet emission of pentacene in the gas phase are not available, this work also presents a synthetic spectrum of the S$_5$~$\rightarrow$ S$_0$ transition relevant to the wavelength range of the observations. Calculated using density functional theory and its time-dependent extension, the synthetic spectrum may facilitate the search for pentacene fluorescence in cometary spectra until laboratory measurements are accessible.
\end{abstract}


}]


\artcitid{\#\#\#\#}
\volnum{000}
\year{0000}
\pgrange{1--}
\setcounter{page}{1}
\lp{1}

\section{Introduction}

Polycyclic aromatic hydrocarbon (PAH) molecules represent a major part of cosmic substances and they are present in cometary matter. First, with various degrees of certainty, observers identified fluorescence bands of free phenanthrene (C$_{14}$H$_{10}$), pyrene (C$_{16}$H$_{10}$), and anthracene (C$_{14}$H$_{10}$) molecules in the coma of comet 1P/Halley \citep[][respectively]{Moreels94,Clairemidi04,Clairemidi08}. Then, leaving no room for doubt as to the presence of PAH substances in comets, mass spectrometry of organic matter from grains collected by space probe Stardust in the coma of comet 81P/Wild~2 revealed, among others, PAH molecules naphthalene (C$_{10}$H$_8$), phenanthrene, and pyrene, as well as several of their alkylated derivatives \citep{Sandford06,Clemett10}. Later, on-site measurements by mass spectrometer ROSINA on board the Rosetta probe evidenced benzene and toluene in the gas of comet 67P/Churyumov-Gerasimenko \citep{Schuhmann19}.

Recently, \citet{Venkataraman23} announced the detection of pentacene (C$_{22}$H$_{14}$) and toluene (C$_7$H$_8$) in the gas of more than a dozen comets after identifying fluorescence bands of the two aromatic substances in the ultraviolet (UV) emission spectra of the objects. The identification of pentacene and toluene bands in the spectra is, however, unsubstantiated when considering laboratory measurements published in the literature. The detection of Fe~{\small{II}} in the comae of the same comets except one is also uncorroborated.

\section{Observations}\label{sec:obs}

\citet{Venkataraman23} proposed a new analysis of cometary spectra measured with the International Ultraviolet Explorer (IUE), no longer in operation, and the Hubble Space Telescope (HST). Together, the Short-Wavelength Spectrograph (SWS) and Long-Wavelength Spectrograph (LWS) of IUE covered the 1150--3200~{\AA} range \citep{Boggess78}. Equipping the HST and both currently operational, the Cosmic Origins Spectrograph (COS) and the Space Telescope Imaging Spectrograph (STIS) cover the 900--3200~{\AA} \citep{McCandliss10,Green12,Hirschauer24} and 1150--10300~{\AA} \citep{Medallon23} ranges, respectively.

The above-mentioned instruments allowed researchers to analyse the light produced by comets, in particular the sunlight-induced UV fluorescence of their gas. Thus, studies clearly identified lines and bands of several atomic and molecular species. For example, \citet{Feldman80} and \citet{Weaver81} identified emission features of H~{\small{I}}, C~{\small{I}}, O~{\small{I}}, S~{\small{I}}, OH, CO, C$_2$, S$_2$, CS, and CO$_2^+$ in the spectra of comets Bradfield 1979l and Meier 1980q, alternatively denoted C/1979 Y1 and C/1980 V1, respectively, in the current style.

\begin{figure*}
\centering\includegraphics[height=.25\textheight]{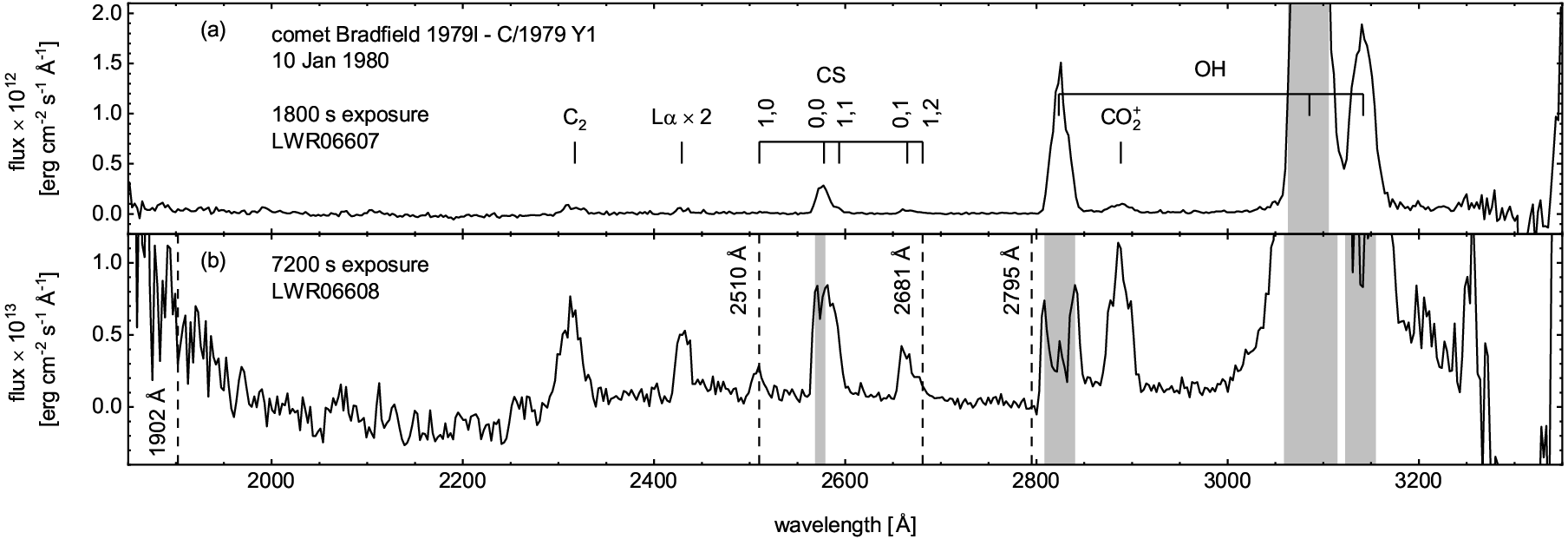}
\caption{Emission spectra LWR06607 and LWR06608 of comet Bradfield 1979l obtained on 10 January 1980 with the long-wavelength redundant (LWR) camera of IUE/LWS. Bradfield 1979l was then at 0.711~AU from the Sun \citep{AHearn80}. The vertical grey bands indicate saturated pixels. (a) Spectrum LWR06607, shorter exposure time. Assignment according to \citet{Feldman80} and \citet{AHearn80}, except for the (1,1) and (1,2) bands of CS A--X (this work, Section~\ref{sec:disc-tolu}). (b) Spectrum LWR06608, longer exposure time. The vertical dashed lines indicate the 1902, 2510, 2681, and 2795~$\AA$ wavelength positions reported by \citet{Venkataraman23}.}\label{fig:Brdfld-LWR}
\end{figure*}

\begin{figure}
\centering\includegraphics[height=.25\textheight]{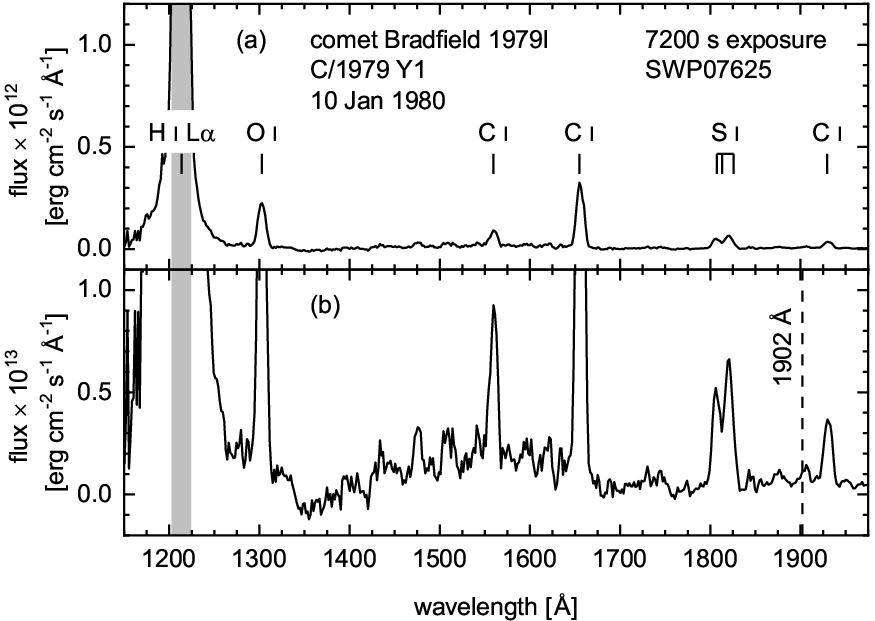}
\caption{Emission spectrum SWP07625 of comet Bradfield 1979l on 10 January 1980 obtained with the short-wavelength prime (SWP) camera of IUE/SWS. Bradfield 1979l was then at 0.711~AU from the Sun \citep{AHearn80}. The vertical grey band signals saturated pixels. (a) Assignment according to \citet{Feldman80}. (b) Vertically magnified view. Bands of the Fourth Positive system of CO, unmarked here, appear weakly between 1400 and 1650~{\AA} \citep{Feldman80,Tozzi98}. The vertical dashed line indicates the 1902~$\AA$ wavelength position reported by \citet{Venkataraman23}.}\label{fig:Brdfld-SWP}
\end{figure}

To illustrate such an analysis, Figures~\ref{fig:Brdfld-LWR} and \ref{fig:Brdfld-SWP} show emission spectra of comet Bradfield 1979l measured with the long- and short-wavelength spectrometers of IUE within a few hours on the same day, after which the emission flux decreased. \citet{Feldman80} and \citet{AHearn80} identified all clear lines and bands.

\begin{figure}
\centering\includegraphics[height=.25\textheight]{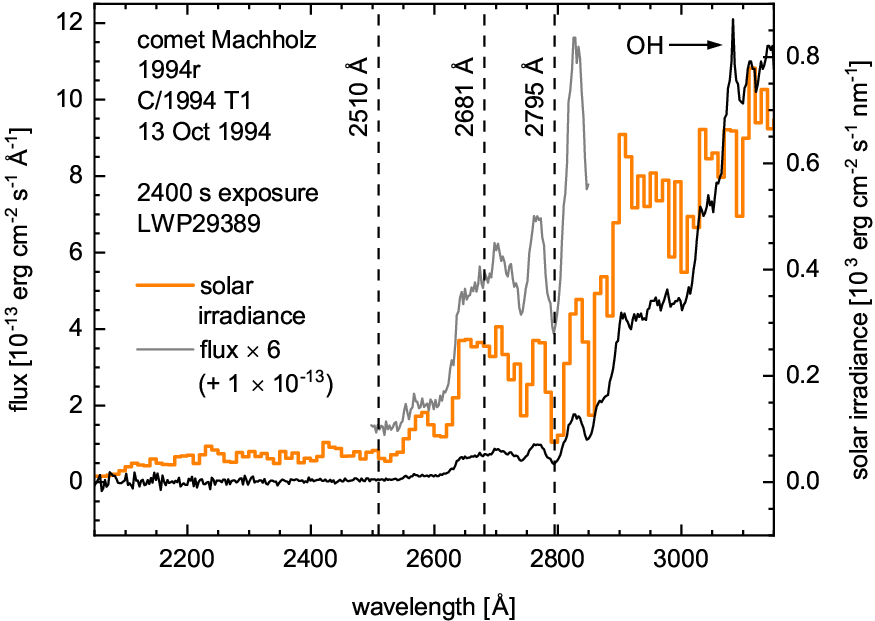}
\caption{Emission spectrum LWP29389 of comet Machholz 1994r on 13 October 1994, Julian date 2449638.93837, obtained with the long-wavelength prime (LWP) camera of IUE/LWS (black curve). Machholz 1994r was then at 1.842~AU from the Sun \citep[extrapolated from data in][]{IAUC6093}. An interval of the spectrum is vertically magnified and offset (grey curve).
The solar irradiance at 1~AU measured almost at the same time on Julian date 2449639.0 (orange curve). The vertical dashed lines indicate the 2510, 2681, and 2795~$\AA$ wavelength positions reported by \citet{Venkataraman23}. The arrow indicates the OH A--X~(0,0) band clearly rising on top of the reflected sunlight.}\label{fig:Mchhlz}
\end{figure}

As comets release both gas and dust, the light that reveals their comae comprises not only the fluorescence of free atoms and molecules, but also the sunlight reflected and scattered by dust grains. The relative strengths of the gas fluorescence and reflected sunlight depend on parameters such as the comet composition and its position at the time of the observation. The solar irradiance spectrum being known \citep[see, for instance,][]{Mount81,Woods96}, it is possible to separate the two contributions. Thus, \citet{Feldman97} subtracted the reflected sunlight from the UV spectrum of comet Hale-Bopp (C/1995 O1).
Figure~\ref{fig:Mchhlz} compares a spectrum of another comet, Machholz 1994r (C/1994 T1),\footnote{\citet{Venkataraman23} mistakenly attributed data LWP29389 to Machholz 1986e (96P/1986 J2).}
with the solar irradiance of the time and shows that the cometary spectrum consists mostly of reflected sunlight. Indeed, the only feature most likely attributable to gas fluorescence is the narrow peak at 3085~{\AA} that coincides with the OH A$^2\Sigma^+$~$\rightarrow$ X$^2\Pi$~(0,0) band, saturated in Figure~\ref{fig:Brdfld-LWR}.

Re-analysing the data that gave Figure~\ref{fig:Brdfld-LWR}a, \citet{Venkataraman23} announced the detection of emission bands at 2681 and 2795~{\AA}, respectively. They also detected the bands in the LWP29389 spectrum of comet Machholz 1994r shown in Figure~\ref{fig:Mchhlz}. In fact, they announced the detection of the two emissions, accompanied by a third one at 1902~{\AA}, in the spectra of more than a dozen comets. Having assigned the bands at 1902 and 2795~{\AA} to pentacene and the one at 2681~{\AA} to toluene, they claimed the detection of the two substances in the cometary comae, presumably as free molecules given that some of the spectra do not show the presence of dust. Additionally, \citet{Venkataraman23} reported the observation of an emission at 2510~{\AA} that they identified as an Fe~{\small{II}} line. The following sections assess the detection and the assignment of the four emissions.

\section{The case of pentacene}

\subsection{Spectroscopy in the visible and UV wavelength domains}\label{sec:spec-pent}

\citet{Venkataraman23} reported emission bands of cometary gas at 1902 and 2795~{\AA} that they attributed to pentacene by relying on studies by \citet{Malloci04} and \citet{Hendrix16}. The first study, however, proposes a theoretical absorption spectrum of pentacene over the 0--30~eV energy range that consists of bands centred at computed vertical electronic transition energies \citep[see Figure~14 in][]{Malloci04}.
It does not include information on vibrational levels, in particular the zero-point energies that contribute to the transition energies in the case of origin bands. Furthermore, its theoretical character means that the accuracy of the calculated electronic energies and derived transition energies depends on the quality of the model chemistry. Thus, in their Theoretical Spectral Database of Polycyclic Aromatic Hydrocarbons, \citet{Malloci07} give, using two models, energies of 3.93 and 4.24~eV for the electronic transition which would correspond to the emission band found by \citet{Venkataraman23} at 2795~{\AA}. The wavelengths derived from the two energies are 3155 and 2924~{\AA}, respectively, and it is not possible to compare them to 2795~{\AA}, if only because the presence of pentacene in cometary gas in detectable quantities remains to be proven.

Theoretical spectra, although useful for interpreting the spectra of identified substances, cannot serve as spectroscopic references for the identification of molecules in a medium of unknown composition, whether it is an interstellar molecular cloud or a cometary coma. Only spectra measured in the laboratory in relevant conditions can serve as references for this purpose. To quote \citet{Malloci04} on the analysis of astronomical spectra in the near-IR, visible, and near-UV window, 'the accuracy of about 0.3~eV achieved by current TD-DFT methods on the position of the bands cannot be used alone for a firm spectral identification'. Although the statement is not recent, it concerns the results to which \citet{Venkataraman23} referred their readers.

The second study presents a reflectance spectrum for a hypothetical surface with the spectral properties of pentacene that \citet{Hendrix16} derived from the synthetic photo-absorption cross-section spectrum computed by \citet{Malloci04} or from similar data available in the Theoretical Spectral Database of Polycyclic Aromatic Hydrocarbons \citep{Malloci07}. Accordingly, it carries the same information with the same accuracy, though under a different shape \citep[see Figure~7 in][]{Hendrix16}, a maximum in absorption giving a minimum in reflectance and vice versa. Derived from a theoretical calculation, this synthetic reflectance spectrum of a hypothetical surface cannot serve as reference for spectroscopic identification. Beyond the scope of this study, reflectance spectra are useful for characterising cometary dust grains through the analysis of the light they reflect.

The absorption spectrum of pentacene does feature a strong band near 2800~{\AA} and it corresponds to the . To date, however, spectroscopic studies of the species in the gas phase have not characterized this band. Nevertheless, absorption spectra of pentacene molecules isolated in rare-gas matrices are available for the UV and visible wavelength domains \citep{Halasinski00} and we can use them to extrapolate approximate band positions for molecules in the gas phase \citep[][and references therein]{Gredel11}. For example, the origin band of the S$_1$~$\leftarrow$ S$_0$ transition of pentacene, for which measurements on molecules in the gas phase are actually available \citep{Amirav80,Amirav81b,Griffiths82,Heinecke98},\footnote{Origin band positions in \citet{Amirav80} and \citet{Amirav81b} strongly differ from those published in other studies, the use of which is to be preferred.}
arises at 542.7 and 559.8~nm when the molecules are isolated in Ne and Ar matrices, respectively \citep{Halasinski00}.\footnote{In agreement, \citet{Szczepanski95} reported 559.7~nm for the measurement on pentacene in an Ar matrix.}
Extrapolation from these values gives a wavelength of 537.5~nm for free pentacene molecules and accurate measurement of the transition in the gas phase indicates a band centre at a wavenumber of 18\,648.996(4)~cm$^{-1}$ \citep{Heinecke98},\footnote{Measured with a vacuum wavemeter.}
to which corresponds a wavelength of 5362.2~{\AA} in vacuum. Thus, the extrapolated wavelength is close to the measured value, though 1.3~nm longer. Application of the procedure to the absorption band of pentacene near 2800~{\AA}, which appears at 281.5 and 291.2~nm in the spectra of molecules isolated in Ne and Ar matrices, respectively \citep{Halasinski00}, gives a wavelength of 278.54~nm for free molecules. Taking the case of the S$_1$~$\leftarrow$ S$_0$ absorption into account, the actual wavelength is possibly slightly shorter.

Following the new convention, UV wavelengths expressed in units of {\AA} in this study refer to calibration in vacuum. Laboratory studies, however, do not consistently indicate the medium, vacuum or air, in which the wavelengths are valid and, whatever the case, their values appear here in units of nm. Nor do they always provide information on the accuracy of the measurements. Nevertheless, an error amounting to 0.5~units of the last given digit is reasonable and deviations caused by an ignored refractive index, 0 for vacuum and $\sim$1.0003 for air at UV wavelengths \citep{Peck72}, are somewhat smaller than 0.1~nm in this region. Therefore, using the extreme values defined by 281.5~$\pm$ 0.05~nm and 291.2~$\pm$ 0.05~nm for measurements in Ne and Ar matrices, the band position for free molecules is 278.54~$\pm$ 0.08~nm when neglecting any correction for a possibly inadequate refractive index. Taking this into account, the extrapolated position becomes 2785.8~$\pm$ 1.2~{\AA}. The extrapolated value is therefore close to that taken into account by \citet{Venkataraman23}, 2795~{\AA}, although shorter by $\sim$9~{\AA}.

To date, the literature does not contain data on the spectrum of free pentacene molecules at and around 1900~{\AA}. Measurements on the molecule isolated in rare-gas matrices or other relevant media do not cover this region either. Consequently, there is insufficient information to plausibly assign a cometary emission band near 1900~{\AA} to this species.

\subsection{Theoretical spectra}

Theoretical calculations on the pentacene molecule indicate that S$_5$(2$^1$B$_{2\mathrm{u}}$) is the upper state of the strong absorption that occurs near 2800~{\AA}. \ref{apx:theo} gives detailed information on the theoretical quantum chemistry model and calculation method.

\begin{figure*}
\centering\includegraphics[height=.25\textheight]{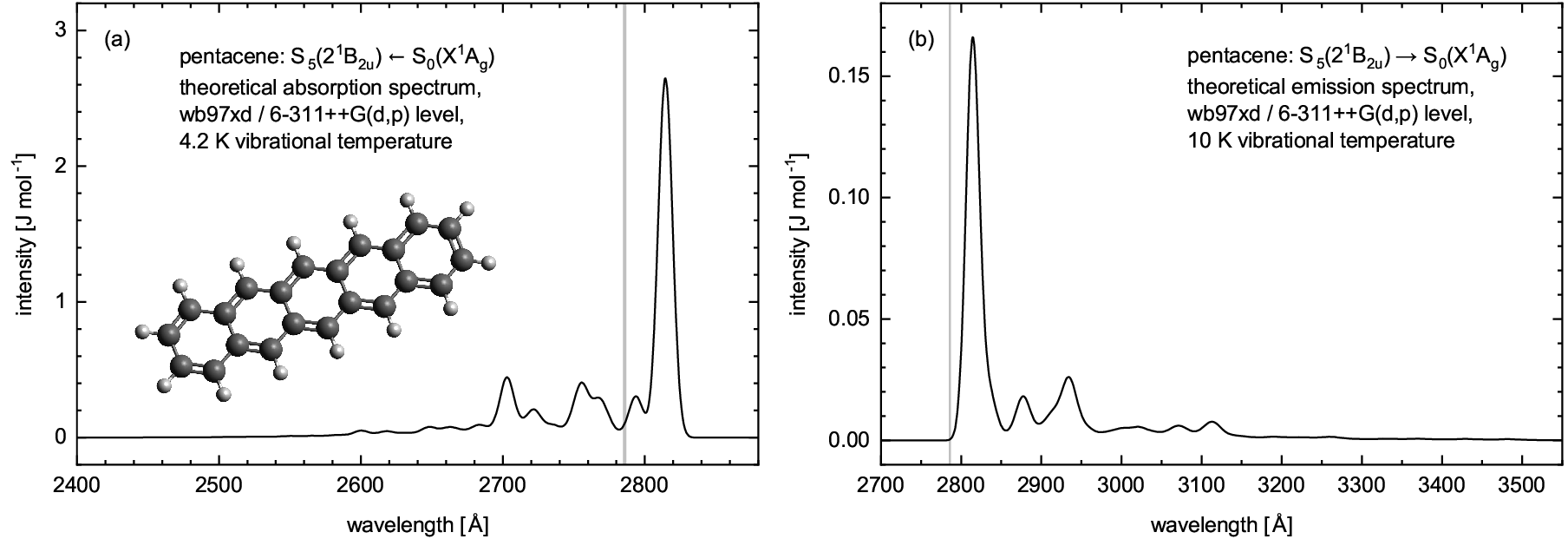}
\caption{Synthetic spectra of vibronic transitions between the S$_0$(X$^1$A$_{\mathrm{g}}$) and S$_5$(2$^1$B$_{2\mathrm{u}}$) states of free pentacene. (a) Absorption spectrum at 4.2~K with a ball-and-stick model representing the structure of pentacene in the electronic ground state S$_0$. The vertical grey narrow band highlights the 2785.8~$\pm$ 1.2~{\AA} interval in which the origin band should arise according to extrapolation from measurements on pentacene isolated in rare-gas matrices (see main text). (b) Emission spectrum of free pentacene at 10~K. The vertical grey narrow band is the same as in panel (a).}\label{fig:theo}
\end{figure*}

The comparison of a synthetic spectrum, derived from theoretical calculations, with a measured spectrum can validate the theoretical model. Figure~\ref{fig:theo}a presents a synthetic S$_5$(2$^1$B$_{2\mathrm{u}}$)~$\leftarrow$ S$_0$(X$^1$A$_{\mathrm{g}}$) absorption spectrum of pentacene for a vibrational temperature of 4.2~K, the temperature that \citet{Halasinski00} set for their study with matrix isolation spectroscopy. The theoretical calculation takes into account up to four quanta of vibrational excitation although the result is not significantly different from taking into account three quanta only. Every band has a Gaussian profile with a full width at half maximum (FWHM) of 150~cm$^{-1}$, similar to the width of the strongest peak in the absorption spectrum of pentacene isolated in Ne ice \citep[FWHM estimated from Figure~2 in][]{Halasinski00}. In the synthetic spectrum, the strongest band represents the (0,0) vibration-less absorption and its wavenumber is 35\,529.18~cm$^{-1}$ for an ideally isolated molecule. Considering the accuracy of theoretical calculations \citep{Malloci04}, the corresponding wavelength of 2814.588~{\AA} is in good agreement with the wavelength extrapolated for molecules in the gas phase, 2785.8~$\pm$ 1.2~{\AA}. The synthetic spectrum strongly resembles the spectrum of Ne-matrix-isolated pentacene and thus the chemistry model appears adapted to calculating spectra between S$_0$ and S$_5$.

Figure~\ref{fig:theo}b shows a synthetic S$_5$(2$^1$B$_{2\mathrm{u}}$)~$\rightarrow$ S$_0$(X$^1$A$_{\mathrm{g}}$) emission spectrum of pentacene for a temperature of 10~K, that is, the temperature at which \citet{Clairemidi04} measured the emission spectrum of pyrene that served as a reference in their successful search of the molecule in the gas of comet 1P/Halley (Halley 1982i, 1/P 1982 U1). The chosen temperature is not critical as the theoretical spectrum does not depend on it significantly and, for example, has the same appearance at 10 and 100~K (\ref{apx:theo}). Although molecules in the gas phase can rotate, the band profiles are not the result of calculations using theoretical rotational constants. For simplicity, all bands have a Gaussian profile with a unique FWHM of 250~cm$^{-1}$, a value approaching that observed in the spectra of cometary PAH molecules \citep{Moreels94,Clairemidi04,Clairemidi08}. Like the synthetic absorption spectrum, this emission spectrum takes into account up to four quanta of vibrational excitation. Thus, assuming that gases released by comets are thermalised --the temperature is not critical-- and that the de-excitation of pentacene molecules in the S$_5$ state proceeds mainly by fluorescence toward S$_0$, the spectrum simulates the emission bands near 2800~{\AA} of pentacene in cometary gas.

As a note, since the synthetic spectrum exhibits only one predominant band, identification of the pentacene emission bands in a cometary spectrum near 2800~{\AA}, where bands of other species such as OH and CO$_2^+$ \citep{Feldman80}, even S$_2$ \citep{AHearn83}, arise, should prove challenging.

\subsection{Cometary spectra and pentacene fluorescence bands}\label{sec:disc-pent}

\begin{figure}
\centering\includegraphics[height=.25\textheight]{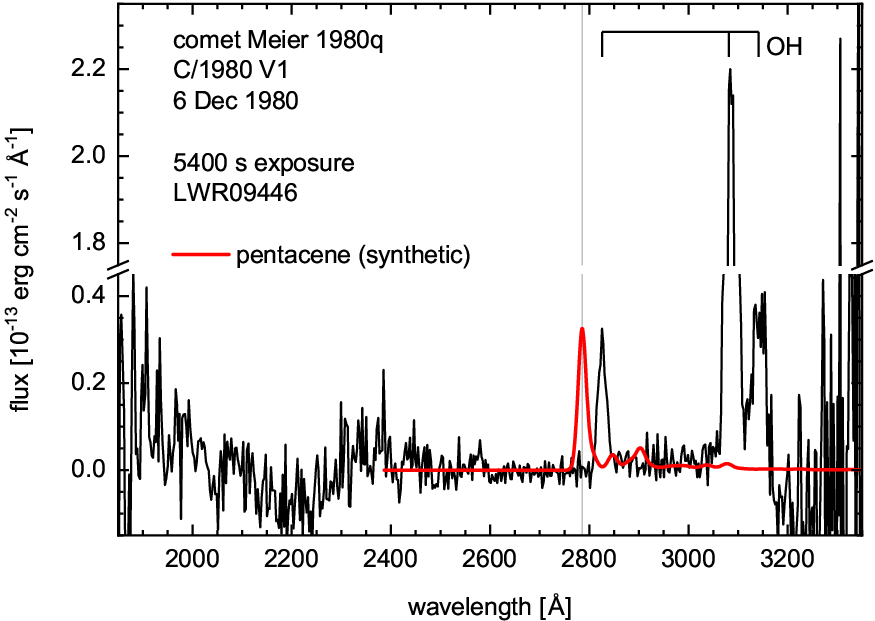}
\caption{Emission spectrum LWR09446 of comet Meier 1980q on 6 December 1980 obtained with the LWR camera of IUE/LWS. Assignment according to \citet{Weaver81}. Meier 1980q was then at 1.52~AU from the Sun \citep{Weaver81}. The red curve is a synthetic S$_5$(2$^1$B$_{2\mathrm{u}}$)~$\rightarrow$ S$_0$(X$^1$A$_{\mathrm{g}}$) emission spectrum of pentacene at 10~K vibrational temperature with a FWHM of 250~cm$^{-1}$ and a peak intensity normalized to that of the nearest OH band. The original synthetic spectrum of pentacene in Figure~\ref{fig:theo}b is shifted in energy by such an amount that the (0,0) band is at 2785.8~{\AA}. The vertical grey narrow band is explained in the caption of Figure~\ref{fig:theo}.}\label{fig:Mr-TDDFT}
\end{figure}

\citet{Venkataraman23} reported the detection of a band at 2795~{\AA} in the spectra of 17 comets and attributed this band to pentacene molecules in the cometary gas. Yet, the close examination of the simplest and clearest spectra plotted in their study, that is, those of comets C/1979~L (C/1979~Y1 or Bradfield 1979l) and C/1980~V1 (Meier 1980q), does not reveal any band at this wavelength \citep[see Figure~2 in][]{Venkataraman23}. \citet{Feldman80} and \citet{Weaver81} published and analysed spectra derived from the very same observational data and these do not show any band at 2795~{\AA} either. Figures~\ref{fig:Brdfld-LWR} and \ref{fig:Mr-TDDFT} show spectra obtained with the same measurements. Though differently processed, they, too, do not show any clear band at 2795~{\AA} or at 2785.8~{\AA}. Consequently, the spectra cannot demonstrate or even suggest the presence of pentacene in the gas of these comets.

As studies revealed fluorescence bands of PAH species in the spectrum of comet 1P/Halley (1P/1982 U1) \citep{Moreels94,Clairemidi04,Clairemidi08}, a new examination of the spectrum might uncover bands of pentacene. \citet{Feldman87} published the UV spectrum of the comet as measured on 11 March 1986, spectrum that actually showed a possible emission band at 2770~{\AA} \citep[see Figure~2 in][]{Feldman87}. Analysis, however, demonstrated that it was in fact a component of the solar irradiance spectrum, shaped by Fraunhofer lines, and hence sunlight reflected by cometary dust. It is clearly visible in the solar irradiance spectrum of Figure~\ref{fig:Mchhlz}, which brings us back to the UV spectra presented by \citet{Venkataraman23}.

\begin{figure}
\centering\includegraphics[height=.25\textheight]{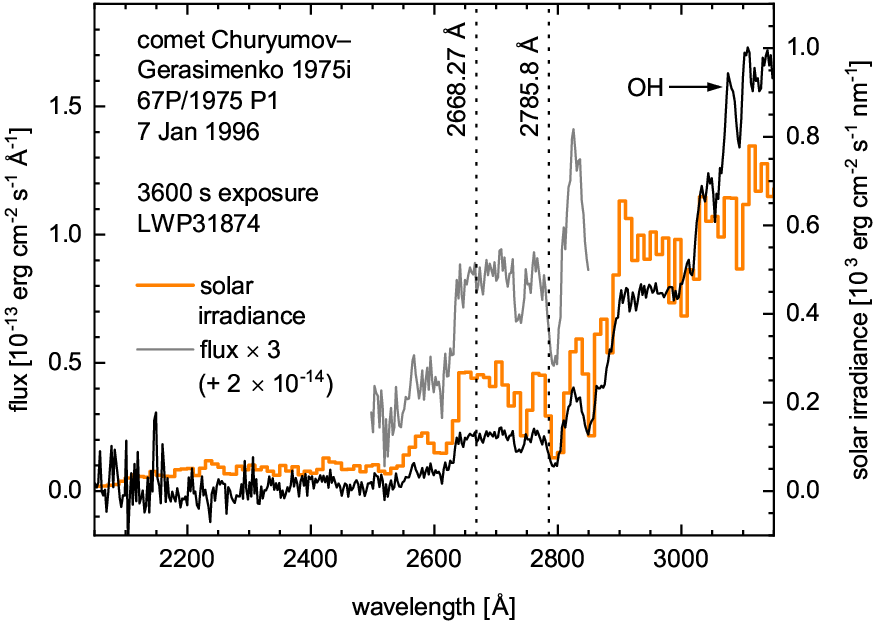}
\caption{Emission spectrum LWP31874 of comet 67P/Churyumov-Gerasimenko on 7 January 1996, Julian date 2450090.34010, obtained with the LWP camera of IUE/LWS (black curve). The comet was then at 1.31~AU from the Sun \citep[read from Figure~10 in][]{MazzottaEpifani09}. An interval of the spectrum is vertically magnified and offset (grey curve). The solar irradiance at 1~AU measured almost at the same time on Julian date 2450090.0 (orange curve). The vertical dotted lines indicate the 2668.27 and 2785.8~$\AA$ wavelength positions of the laboratory-measured S$_1$--S$_0$ (0,0) band of toluene and the extrapolated S$_5$--S$_0$ (0,0) band of pentacene, respectively. The arrow indicates the OH A--X~(0,0) band rising on top of the reflected sunlight.}\label{fig:Chury}
\end{figure}

Among them, the spectrum of comet Machholz 1994r (C/1994 T1) consists mostly of solar light reflected by cometary dust grains as remarked in Section~\ref{sec:obs}. Figure~\ref{fig:Mchhlz}, using the data exploited by \citet{Venkataraman23}, illustrates the case. A narrow peak at 3085~{\AA} that coincides with the OH A--X~(0,0) band appears as the only likely sign of gas fluorescence. Similarly, in Figure~\ref{fig:Chury}, the spectrum of comet 67P/Churyumov–Gerasimenko (67P/1975 P1 or 1975i) clearly exhibits the characteristic pattern of sunlight between 2600 and 2850~{\AA}. The strong solar contribution greatly hampers the search for pentacene fluorescence in the emission spectrum of this comet. In fact, the spectrum does not show clear signs of a band at 2795~{\AA} or at 2785.8~{\AA}, although the data are those analysed by \citet{Venkataraman23}.

\begin{figure}
\centering\includegraphics[height=.25\textheight]{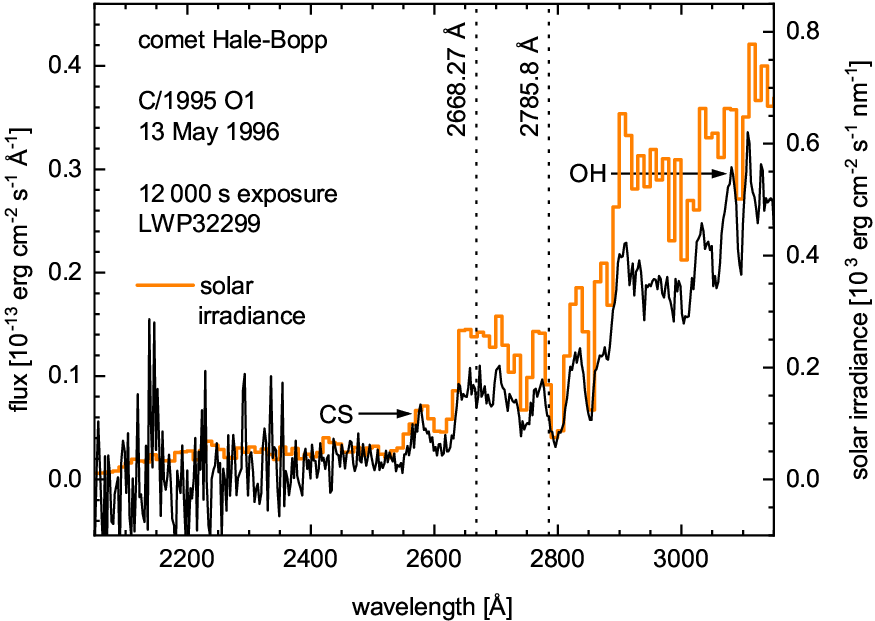}
\caption{Emission spectrum LWP32299 of comet Hale-Bopp on 13 May 1996, Julian date 2450216.63779, obtained with the LWP camera of IUE/LWS (black curve). Hale-Bopp was then at 4.43~AU from the Sun \citep{Weaver97}. The solar irradiance at 1~AU measured almost at the same time on Julian date 2450217.0 (orange curve). The vertical dotted lines indicate the 2668.27 and 2785.8~$\AA$ wavelength positions of the laboratory-measured S$_1$--S$_0$ (0,0) band of toluene and the extrapolated S$_5$--S$_0$ (0,0) band of pentacene, respectively. Arrows indicate the CS A--X~(0,0) and OH A--X~(0,0) bands peaking weakly on top of the reflected sunlight.}\label{fig:HaleBopp}
\end{figure}

Figure~\ref{fig:HaleBopp} shows another example of major solar contribution with a spectrum of comet Hale-Bopp (C/1995 O1). \citet{Venkataraman23} revisited the same spectrum LWP32299 measured with IUE/LWS on 13 May 1996 as the comet was at a heliocentric distance of 4.43~AU. Its outline is similar to that of a spectrum (likely Y3DU0605T) measured later on 18 October 1996 with the Faint Object Spectrograph (FOS) as the heliocentric distance was 2.69~AU. Although the production rate of gas had increased in a greater proportion than that of dust between the two dates, both spectra and those measured in the interval did not reveal more than the CS A--X~(0,0) and OH A--X~(0,0) bands, that is, after subtraction of the reflected sunlight \citep{Feldman97}. The observations with HST/FOS additionally featured the (1,0) and (0,0) CO Cameron bands near 2000~{\AA} (a$^3\Pi$--X$^1\Sigma^+$ system) from late June to mid-October. \citet{Weaver97} and \citet{Feldman97} reported the preceding information on observations of comet Hale-Bopp with IUE and HST.

Thus, the spectra of comets Machholz 1994r, 67P/Churyumov-Gerasimenko, and Hale-Bopp do not prove the presence of the acene in the corresponding comets. The same observation and conclusion possibly apply to the remaining spectra measured with IUE as well as all those measured with the HST \citep[Figures~2 and 4 in][]{Venkataraman23}.

Concerning the emission band that \citet{Venkataraman23} saw at 1902~{\AA} in the spectra of 21 comets, it is very weak in the spectrum of comet Bradfield 1979l displayed in Figure~\ref{fig:Brdfld-SWP} \citep[see also Figure~3 in][]{Venkataraman23}. Anyway, as mentioned in Section~\ref{sec:spec-pent}, there is not enough information about the spectrum of pentacene around this wavelength to convincingly assign the band to this molecule.

\section{The case of toluene}\label{sec:disc-tolu}

\citet{Venkataraman23} found a band at 2681~{\AA} in their selection of cometary spectra and assigned it to toluene with reference to \citet{Sunuwar21}. The reference reports among others the absorption and emission spectra of toluene in $n$-hexane solution at room temperature. In Figure~\ref{fig:Brdfld-LWR}b, the spectrum of comet Bradfield 1979l actually shows a possible band at this wavelength in the form of a shoulder on the red side of the A($^1\Pi$)~$\rightarrow$ X($^1\Sigma^+$)~(0,1) band of CS. The strongest (0,0) band exhibits a similar shoulder (Figure~\ref{fig:Brdfld-LWR}a). Given that the (1,0) band is visible, the shoulders of the (0,0) and (0,1) bands are most likely the (1,1) and (1,2) bands, respectively. This assignment is consistent with the band positions measured in the laboratory by \citet{Bergeman81}, that is, 39\,832.9~cm$^{-1}$ for the (1,0) band of CS A--X, 38\,797.1 and 38\,560.8~cm$^{-1}$ for the (0,0) and (1,1) bands, and 37\,524.9 and 37\,301.4~cm$^{-1}$ for the (0,1) and (1,2) bands. The corresponding wavelengths in vacuum are 2510.49, 2577.51, 2593.31, 2664.90, and 2680.86~{\AA}, respectively. The bands appeared resolved in a spectrum of comet Shoemaker-Levy C/1991 T2 obtained with HST/FOS (spectrum Y0YZ1802T) and \citet{Feldman96} attributed the bands to CS in one of the spectra measured at the time \citep[see Figure~4 in][]{Feldman96}.

Still, the S$_1$--S$_0$ system of toluene has its origin band near 2681~{\AA}. \citet{Ginsburg46} measured the (0,0) absorption band at 37\,477.4~cm$^{-1}$ in vacuum, equivalent to 2668.27~{\AA}, also in vacuum. Later, \citet{Selco89} reported a close value of 37\,474~cm$^{-1}$ in vacuum with dispersed emission spectroscopy. Thus, the S$_1$~$\rightarrow$ S$_0$ (0,0) band of toluene is 13~{\AA} to the blue of the cometary band that \citet{Venkataraman23} assigned to the molecule. Incidentally, its position is between the overlapping A($^1\Pi$)~$\rightarrow$ X($^1\Sigma^+$)~(0,1) and (1,2) bands of CS, closer to the former. As a result, the fluorescence of CS hinders the observation of the S$_1$~$\rightarrow$ S$_0$ (0,0) band of toluene, the presence of which is not clear. Nevertheless, photo-excitation of toluene through S$_1$~$\leftarrow$ S$_0$ induces an emission spectrum that comprises numerous bands reflecting the vibrational structure of the electronic ground state, in particular between 2720 and 2750~{\AA} \citep[see, for instance, measurements at room temperature by][]{Blondeau71}. Additionally, a photo-excitation source with a broad-band spectrum such as sunlight would also cause a broad emission component over the 2700--3000~{\AA} range \citep{Blondeau71}. There is no sign of such a pattern or broad emission in the spectrum of comet Bradfield 1979l, however, despite the favourable condition of the absence of dust reflecting sunlight (Figure~\ref{fig:Brdfld-LWR}).

Interestingly, the exploration of comet 67P/Churyumov-Gerasimenko (67P/1975 P1) with the Rosetta spacecraft equipped with the mass spectrometer ROSINA evidenced the presence of toluene in the gas of its coma \citep{Schuhmann19}. As remarked in Section~\ref{sec:disc-pent}, the spectrum that \citet{Venkataraman23} presented consists essentially of sunlight reflected by dust grains, the only contribution from the gas being the OH A--X~(0,0) band that possibly appears on top of it at 3085~{\AA}. Figure~\ref{fig:Chury} shows the relevant spectra and the S$_1$~$\rightarrow$ S$_0$ (0,0) band of toluene is not apparent at 2668.27~{\AA}. As to a possible broad emission by toluene between 2700--3000~{\AA}, the reflected sunlight prevents the detection of anything weak. To be noted, the abundance of toluene relative to water in the cometary gas was (6.16~$\pm$ 1.23)~$\times$ 10$^{-5}$ in May 2015 as the comet was at 1.52~AU from the Sun \citep{Schuhmann19}. The heliocentric distance was 1.31~AU for the measurement of the spectrum displayed in Figure~\ref{fig:Chury}, therefore the gas production rate would be higher. Yet, this is a spectrum dominated by reflected sunlight and OH, product of H$_2$O dissociation, is barely perceivable. Thus, although the substance was present and detected in situ with ROSINA, its abundance was not high enough for optical detection with IUE.

\section{The case of Fe~{\small{II}}}\label{sec:disc-feii}

Beside the detection of pentacene and toluene fluorescence bands, \citet{Venkataraman23} claimed the observation of an Fe~{\small{II}} emission line at 2510~{\AA} in 16 spectra of comets out of 17, the exception being the spectrum of Machholz 1994r. All known Fe~{\small{II}} lines between 2500 and 2520~{\AA} correspond to transitions between excited states of this ion, however \citep{NIST_ASD}. While observing emission between excited states is possible, one would expect to also observe emission ending at the ground state when it is in the same wavelength range. For instance, Fe~{\small{II}} shows such emission lines at 2344.2, 2374.5, 2382.8, 2586.6, and 2600.2~{\AA} \citep{NIST_ASD}. According to a spectrum of Earth's ionosphere above 400~km \citep{Dymond03}, those at 2382.8 and 2600.2~{\AA} are prominent while no emission is detected at 2510~{\AA}.

\begin{figure}
\centering\includegraphics[height=.25\textheight]{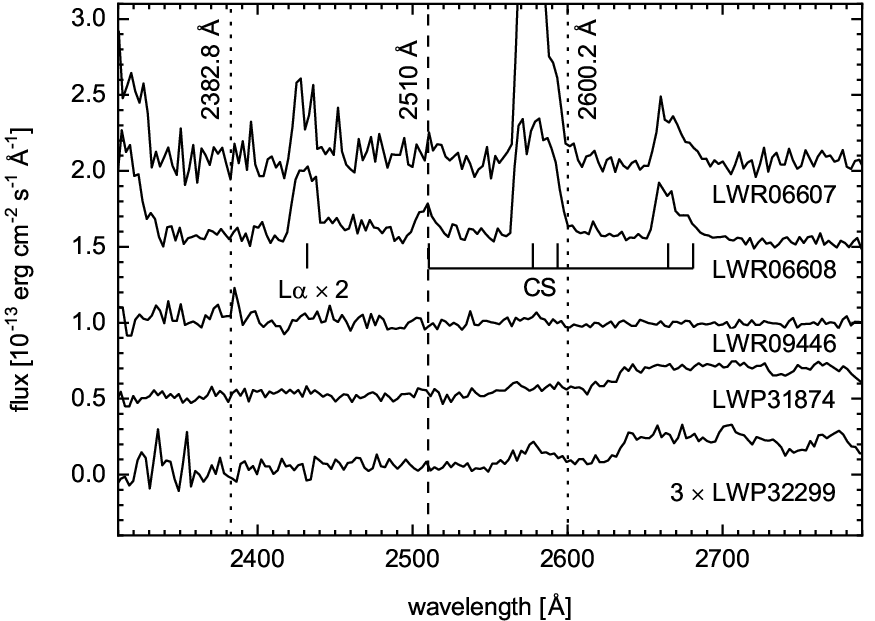}
\caption{From top to bottom, emission spectra of comets Bradfield 1979l (LWR06607 and LWR06608), Meier 1980q (LWR09446), 67P/Churyumov-Gerasimenko (LWP31874), and Hale-Bopp (LWP32299) obtained with the LWR and LWP cameras of IUE/LWS. Labels and marks indicate the Lyman-$\alpha$ emission line and CS emission bands. Dotted vertical lines indicate the wavelength positions of Fe~{\small{II}} emission lines at 2382.8 and 2600.2~{\AA} \citep{NIST_ASD}. The vertical dashed line indicates the wavelength position 2510~{\AA} reported by \citet{Venkataraman23}.}\label{fig:FeII}
\end{figure}

Figure~\ref{fig:FeII} allows for a close examination of spectra in which \citet{Venkataraman23} observed Fe~{\small{II}} at 2510~{\AA}. The spectra of comets Bradfield 1979l, Meier 1908q, 67P/Churyumov-Gerasimenko, and Hale-Bopp, also displayed in Figures~\ref{fig:Brdfld-LWR}, \ref{fig:Mr-TDDFT}, \ref{fig:Chury}, and \ref{fig:HaleBopp}, do not reveal lines attributable to Fe~{\small{II}}, neither at the expected wavelengths of 2382.8 and 2600.2~{\AA} nor at the wavelength of 2510~{\AA} indicated by \citet{Venkataraman23}. Thus, the detection of the ion by \citet{Venkataraman23} in the spectra of comets Bradfield 1979l, Meier 1980q, 67P/Churyumov-Gerasimenko, and Hale-Bopp is not plausible.

Although the present spectra do not show Fe~{\small{II}} emission lines, Fe atoms should be present in the gas of the comets discussed here, for \citet{Manfroid21}  found that Fe~{\small{I}} lines were ubiquitous in the near-UV and visible spectra of cometary atmospheres. Actually, Fe appeared in mass spectra measured in the gas of 67P/Churyumov-Gerasimenko with ROSINA during the Rosetta mission \citep{Rubin22}. Nevertheless, the strong Fe~{\small{I}} 2484.0~{\AA} emission line \citep{NIST_ASD} is not visible in any of the spectra obtained with IUE/LWS examined here, including the spectrum of 67P/Churyumov-Gerasimenko (Figure~\ref{fig:Chury}). Moreover, \citet{Venkataraman23} did not spot this line in the spectra of comets for which \citet{Manfroid21} reported the observation of Fe~{\small{I}} lines at visible wavelengths. It is inferred that the typical parameters and conditions of observations with IUE/LWS and HST/COS were not favourable for detection of fluorescence of Fe atoms, possibly metal atoms in general, at UV wavelengths.

\vspace{-2em}
\section{Conclusion}
\citet{Venkataraman23} announced the detection of pentacene and toluene fluorescence bands in UV spectra of cometary gas. They identified two bands of pentacene, at 1902 and 2795~{\AA}, and one band of toluene, at 2681~{\AA}, in a selection of archival spectra. They also reported the observation of an Fe~{\small{II}} emission line at 2510~{\AA} that would occur between two excited states. Their claims presents issues.

First, information on the UV emission spectrum of pentacene in the gas phase is non-existent. To date, the search for fluorescence bands of this acene has to rely on wavelengths extrapolated from absorption spectra of pentacene isolated in rare-gas matrices and on relative band intensities obtained with theoretical calculations. Accordingly, the origin band of the S$_5$~$\rightarrow$ S$_0$ emission would be at 2785.8~$\pm$ 1.2~{\AA} and would clearly dominate the spectrum. The attribution of an emission band at 2795~{\AA} to pentacene is tentative at best, and that of a band at 1902~{\AA} is not justified. Finally, the cometary spectra selected by \citet{Venkataraman23} do not show clear bands at either wavelength in a consistent manner.

Second, according to laboratory measurements, the S$_1$~$\rightarrow$ S$_0$ fluorescence of toluene has its origin band at 2668.27~{\AA} in vacuum. It overlaps the A($^1\Pi$)~$\rightarrow$ X($^1\Sigma^+$) (0,1) band of CS that hampers its detection when the diatomic is abundant. In any case, the detection of other bands of the S$_1$~$\rightarrow$ S$_0$ spectrum, for example, between 2720 and 2750~{\AA}, is necessary to confirm the detection of the aromatic. The cometary spectra selected by \citet{Venkataraman23} do not show these bands. Incidentally, the band closest to 2681~{\AA} in the spectrum of comet Bradfield 1979l is the A($^1\Pi$)~$\rightarrow$ X($^1\Sigma^+$) (1,2) band of CS. This assignment and that of the (1,1) band of the system are possibly original.

Third, the cometary spectra do not show emission lines of Fe~{\small{II}} at 2382.8 and 2600.2~{\AA}.

To conclude, the examination of the cometary spectra selected by \citet{Venkataraman23} does not reveal emission bands and lines attributable to pentacene, toluene, or Fe~{\small{II}}. Therefore, the claim that they detected the fluorescence of the tree species in these spectra is unsubstantiated. In the case of pentacene, spectra of the S$_5$--S$_0$ system are not available for the molecule in the gas phase. The synthetic spectrum of the S$_5$~$\rightarrow$ S$_0$ transition that this study proposes may help in the search for free pentacene molecules in cometary comae until laboratory measurements are accessible.

\appendix

\section{Theoretical calculations}\label{apx:theo}

The calculation of the electronic structure of pentacene in its ground (S$_0$) and excited (S$_5$) electronic states used density functional theory (DFT) and its time-dependent extension (TDDFT), respectively, as implemented in the Gaussian software \citep{Gaussian16}. It combined the wb97xd functional \citep{Chai08} with the 6-311++g(d,p) basis set \citep{Frisch84}. The optimization of both electronic structures applied the tight criterion of the software and also its ultrafine grid for calculation of integrals. Calculation of their vibrational modes verified that the optimized structures corresponded to minima on the respective potential energy surfaces. It did not take anharmonicity into account, which may be a reasonable choice at least concerning the ground state \citep{Griffiths82}. Nevertheless, \citet{Lemmens19} showed that taking anharmonicity into account improved the theoretical infrared absorption spectra of polyacenes. Therefore, it may also improve theoretical vibronic spectra, in particular concerning their dependence on temperature.

\begin{figure}
\centering\includegraphics[height=.58\textheight]{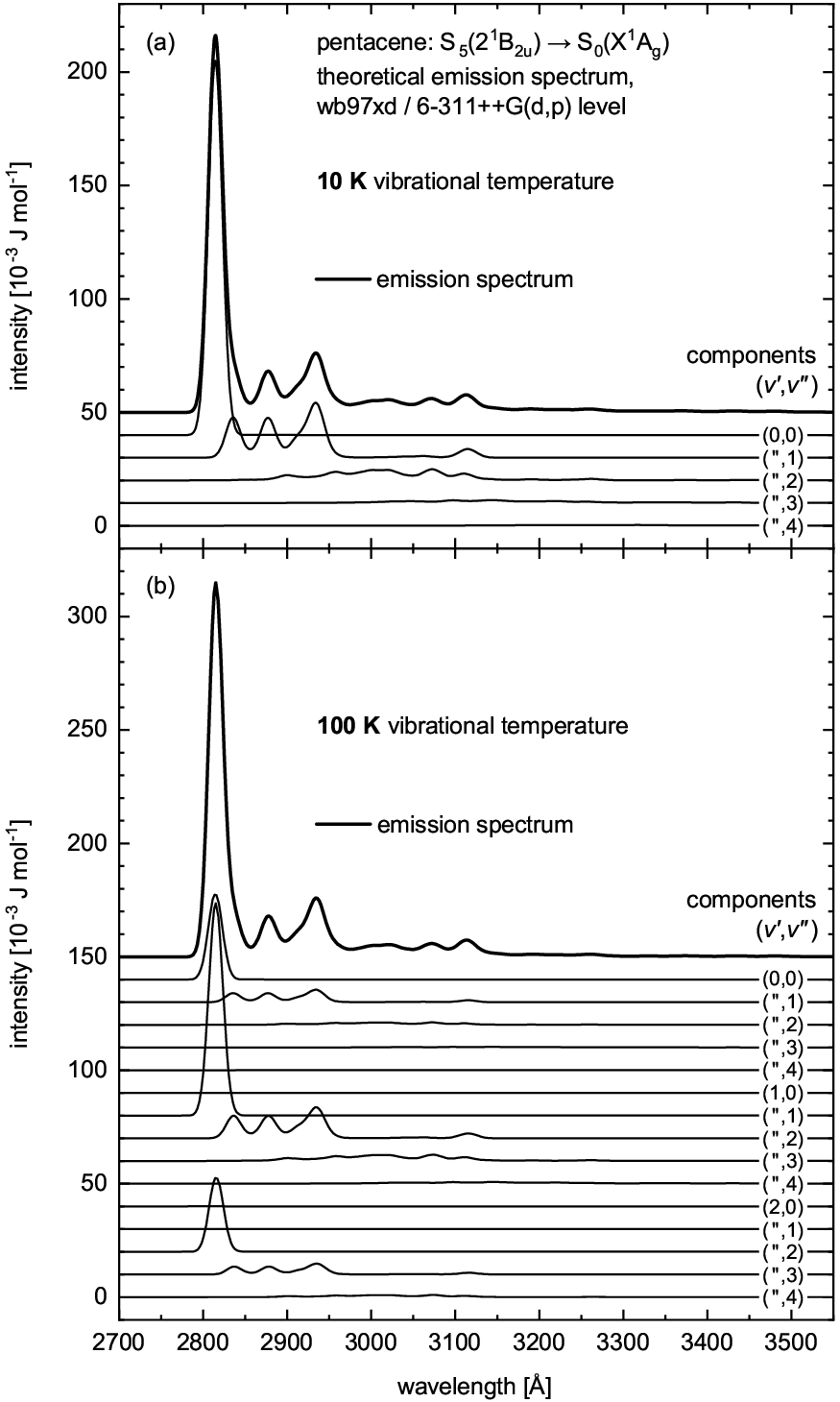}
\caption{Synthetic spectra of the S$_5$(2$^1$B$_{2\mathrm{u}}$)~$\rightarrow$ S$_0$(X$^1$A$_{\mathrm{g}}$) transition of free pentacene and their decomposition into vibrational components. The panels display only components rising above 0.001~J mol$^{-1}$ among all ($v'$,$v''$) combinations, with $v'$ and $v''$ from 0--4. (a) Spectrum at 10~K vibrational temperature. (b) Spectrum at 100~K vibrational temperature.}\label{fig:10K100K}
\end{figure}

Regarding the production of the synthetic absorption and emission spectra displayed in Figure~\ref{fig:theo}, the calculation with the Gaussian software used the calculated theoretical harmonic modes, their frequencies, and vibrational levels with up to four quanta of excitation in total populated according to temperature. It did not implement any scaling of the vibrational frequencies. Concerning the present S$_5$--S$_0$ system of pentacene, the Franck-Condon factors are negligible for ($v'$,$v''$~$<$ $v'$) emission bands. As a result, increasing the temperature does not give rise to noticeable bands on the blue side of the (0,0) emission band. At the same time, the harmonic nature of the vibrational modes gives intensity emission patterns that do not visibly depend on temperature. Simply, the loss of intensity on the ($v'$,$v''$~$\geq$ $v'$) emission is compensated by the gains on the ($v'$~+ $\Delta$$v$,$v''$~+ $\Delta$$v$) emissions that arise at the same wavelength. Figure~\ref{fig:10K100K} illustrates the result of the combined properties.

With the wb97xd/6-311++g(d,p) chemistry model, the excited state that corresponds to absorption near 2800~{\AA} is S$_5$ given the computed vertical electronic transition energies. It is the second singlet electronic state with a wave function that transforms according to the B$_{2\mathrm{u}}$ irreducible representation of the D$_{2\mathrm{h}}$ point group, hence the name S$_5$(2$^1$B$_{2\mathrm{u}}$). \citet{Halasinski00} gave the representation as B$_{3\mathrm{u}}$ because they chose the molecular axes according to the convention set by \citet{Pariser56} while this study applied the convention recommended by \citet{Mulliken55}.

\section*{Acknowledgements}

The author acknowledge the use of data collected with the International Ultraviolet Explorer, specifically within observing programs SCBMA for comet Bradfield 1979l (C/1979 Y1), SCCPF for comet Meier 1980q (C/1980 V1), CORMA for comet Machholz 1994r (C/1994 T1), and SS129 for comets 67P/Churyumov-Gerasimenko 1975i (67P/1975 P1) and Hale-Bopp (C/1995 O1). 
The data were obtained from the Mikulski Archive for Space Telescopes (MAST) at the Space Telescope Science Institute. STScI is operated by the Association of Universities for Research in Astronomy, Inc., under NASA contract NAS5–26555. Support to MAST for these data is provided by the NASA Office of Space Science via grant NAG5–7584 and by other grants and contracts.
The data of the IUE mission were prepared by the NASA Space Science Data Coordinated Archive (NSSDC) and were used in this study without further processing.
The results presented in this document rely on data measured from the Upper Atmosphere Research Satellite (UARS) Solar-Stellar Irradiance Comparison Experiment (SOLSTICE). These data are available from the UARS-SOLSTICE website at https://lasp.colorado.edu/solstice/data/. These data were accessed via the LASP Interactive Solar Irradiance Datacenter (LISIRD) at https://lasp.colorado.edu/lisird/.
The author is grateful to C. J{\"a}ger for access to the Gaussian software.
\vspace{-1em}

\bibliography{pentacene-2}

\begin{thebibliography}{}
\expandafter\ifx\csname natexlab\endcsname\relax\def\natexlab#1{#1}\fi

\bibitem[{{A'Hearn} \& Feldman(1980)}]{AHearn80}
{A'Hearn}, M.~F., \& Feldman, P.~D. 1980, {ApJ}, 242, L187

\bibitem[{{A'Hearn} {$et~al$.}(1983){A'Hearn}, Feldman, \&
  Schleicher}]{AHearn83}
{A'Hearn}, M.~F., Feldman, P.~D., \& Schleicher, D.~G. 1983, {ApJ}, 274, L99

\bibitem[{Amirav {$et~al$.}(1980)Amirav, Even, \& Jortner}]{Amirav80}
Amirav, A., Even, U., \& Jortner, J. 1980, {C}hem. {P}hys. {L}ett., 72, 21

\bibitem[{Amirav {$et~al$.}(1981)Amirav, Even, \& Jortner}]{Amirav81b}
---. 1981, {J}. {P}hys. {C}hem., 85, 309

\bibitem[{Bergeman \& Cossart(1981)}]{Bergeman81}
Bergeman, T., \& Cossart, D. 1981, {J}. {M}ol. {S}pectrosc., 87, 119

\bibitem[{Blondeau \& Stockburger(1971)}]{Blondeau71}
Blondeau, J.~M., \& Stockburger, M. 1971, {B}er. {B}unsenges. {P}hys. {C}hem.,
  75, 450

\bibitem[{Boggess {$et~al$.}(1978)Boggess, Carr, Evans, Fischel, Freeman,
  Fuechsel, Klinglesmith, Krueger, Longanecker, Moore, Pyle, Rebar, Sizemore,
  Sparks, Underhill, Vitagliano, West, Macchetto, Fitton, Barker, Dunford,
  Gondhalekar, Hall, Harrison, Oliver, Sandford, Vaughan, Ward, Anderson,
  Boksenberg, Coleman, Snijders, \& Wilson}]{Boggess78}
Boggess, A., Carr, F.~A., Evans, D.~C., {$et~al$.} 1978, {N}ature, 275, 372

\bibitem[{Chai \& {Head-Gordon}(2008)}]{Chai08}
Chai, J.-D., \& {Head-Gordon}, M. 2008, {P}hys. {C}hem. {C}hem. {P}hys., 10,
  6615

\bibitem[{Clairemidi {$et~al$.}(2004)Clairemidi, Br{\'e}chignac, Moreels, \&
  Pautet}]{Clairemidi04}
Clairemidi, J., Br{\'e}chignac, P., Moreels, G., \& Pautet, D. 2004, {P}lanet.
  {S}pace {S}ci., 52, 761

\bibitem[{Clairemidi {$et~al$.}(2008)Clairemidi, Moreels, Mousis, \&
  Br{\'e}chignac}]{Clairemidi08}
Clairemidi, J., Moreels, G., Mousis, O., \& Br{\'e}chignac, P. 2008, {A}\&{A},
  492, 245

\bibitem[{Clemett {$et~al$.}(2010)Clemett, Sandford, {Nakamura-Messenger},
  H{\"o}rz, , \& {McKay}}]{Clemett10}
Clemett, S.~J., Sandford, S.~A., {Nakamura-Messenger}, K., {$et~al$.} 2010,
  {M}eteorit. {P}lanet. {S}ci., 45, 701

\bibitem[{Dymond {$et~al$.}(2003)Dymond, Wolfram, Budzien, Nicholas, {McCoy},
  \& Thomas}]{Dymond03}
Dymond, K.~F., Wolfram, K.~D., Budzien, S.~A., {$et~al$.} 2003, {G}eophys.
  {R}es. {L}ett., 30, 1003

\bibitem[{Feldman(1996)}]{Feldman96}
Feldman, P.~D. 1996, {C}omets, in Science with the Hubble Space Telescope --
  II, P. Benvenuti, F. D. Macchetto, and E. J. Schreier, eds., Space Telescope
  Science Institute, pp. 525--531

\bibitem[{Feldman(1997)}]{Feldman97}
---. 1997, {E}arth {M}oon {P}lanets, 79, 145

\bibitem[{Feldman {$et~al$.}(1980)Feldman, Weaver, Festou, {A'Hearn}, Jackson,
  Donn, Rahe, Smith, \& Benvenuti}]{Feldman80}
Feldman, P.~D., Weaver, H.~A., Festou, M.~C., {$et~al$.} 1980, {N}ature, 286,
  132

\bibitem[{Feldman {$et~al$.}(1987)Feldman, Festou, {A'Hearn}, Arpigny,
  Butterworth, Cosmovici, Danks, Gilmozzi, Jackson, {McFadden}, Patriarchi,
  Schleicher, Tozzi, Wallis, Weaver, \& Woods}]{Feldman87}
Feldman, P.~D., Festou, M.~C., {A'Hearn}, M.~F., {$et~al$.} 1987, {A}\&{A},
  187, 325

\bibitem[{Frisch {$et~al$.}(1984)Frisch, Pople, \& Binkley}]{Frisch84}
Frisch, M.~J., Pople, J.~A., \& Binkley, J.~S. 1984, {J}. {C}hem. {P}hys., 80,
  3265

\bibitem[{Frisch {$et~al$.}(2016)Frisch, Trucks, Schlegel, Scuseria, Robb,
  Cheeseman, Scalmani, Barone, Petersson, Nakatsuji, Li, Caricato, Marenich,
  Bloino, Janesko, Gomperts, Mennucci, Hratchian, Ortiz, Izmaylov, Sonnenberg,
  Williams-Young, Ding, Lipparini, Egidi, Goings, Peng, Petrone, Henderson,
  Ranasinghe, Zakrzewski, Gao, Rega, Zheng, Liang, Hada, Ehara, Toyota, Fukuda,
  Hasegawa, Ishida, Nakajima, Honda, Kitao, Nakai, Vreven, Throssell,
  {Montgomery, Jr.}, Peralta, Ogliaro, Bearpark, Heyd, Brothers, Kudin,
  Staroverov, Keith, Kobayashi, Normand, Raghavachari, Rendell, Burant,
  Iyengar, Tomasi, Cossi, Millam, Klene, Adamo, Cammi, Ochterski, Martin,
  Morokuma, Farkas, Foresman, \& Fox}]{Gaussian16}
Frisch, M.~J., Trucks, G.~W., Schlegel, H.~B., {$et~al$.} 2016,
  {G{\footnotesize AUSSIAN}}

\bibitem[{Ginsburg {$et~al$.}(1946)Ginsburg, Robertson, \& Matsen}]{Ginsburg46}
Ginsburg, N., Robertson, W.~W., \& Matsen, F.~A. 1946, {J}. {P}hys. {C}hem.,
  14, 511

\bibitem[{Gredel {$et~al$.}(2011)Gredel, Carpentier, Rouill{\'e}, Steglich,
  Huisken, \& Henning}]{Gredel11}
Gredel, R., Carpentier, Y., Rouill{\'e}, G., {$et~al$.} 2011, {A}\&{A}, 530,
  A26

\bibitem[{Green {$et~al$.}(2012)Green, Froning, Osterman, Ebbets, Heap,
  Leitherer, Linsky, Savage, Sembach, Shull, Siegmund, Snow, Spencer, Stern,
  Stocke, Welsh, B{\'e}land, Burgh, Danforth, France, Keeney, {McPhate},
  Penton, Andrews, Brownsberger, Morse, \& Wilkinson}]{Green12}
Green, J.~C., Froning, C.~S., Osterman, S., {$et~al$.} 2012, {ApJ}, 744, 60

\bibitem[{Griffiths \& Freedman(1982)}]{Griffiths82}
Griffiths, A.~M., \& Freedman, P.~A. 1982, {J}. {C}hem. {S}oc. {F}araday
  {T}rans. 2, 78, 391

\bibitem[{Halasinski {$et~al$.}(2000)Halasinski, Hudgins, Salama, Allamandola,
  \& Bally}]{Halasinski00}
Halasinski, T.~M., Hudgins, D.~M., Salama, F., Allamandola, L.~J., \& Bally, T.
  2000, {J}. {P}hys. {C}hem. {A}, 104, 7484

\bibitem[{Heinecke {$et~al$.}(1998)Heinecke, Hartmann, M{\"u}ller, \&
  Hese}]{Heinecke98}
Heinecke, E., Hartmann, D., M{\"u}ller, R., \& Hese, A. 1998, {J}. {C}hem.
  {P}hys., 109, 906

\bibitem[{Hendrix {$et~al$.}(2016)Hendrix, Vilas, \& Li}]{Hendrix16}
Hendrix, A.~R., Vilas, F., \& Li, J.-Y. 2016, {M}eteorit. {P}lanet. {S}ci., 51,
  105

\bibitem[{{Hirschauer, A. S., et al.}(2024)}]{Hirschauer24}
{Hirschauer, A. S., et al.} 2024, {C}osmic {O}rigins {S}pectrograph
  {I}nstrument {H}andbook, {V}ersion 17.0, {B}altimore: {STScI}

\bibitem[{Kramida {$et~al$.}(2024)Kramida, Ralchenko, Reader, \& {NIST ASD
  Team}}]{NIST_ASD}
Kramida, A., Ralchenko, Y., Reader, J., \& {NIST ASD Team}. 2024, {NIST}
  {A}tomic {S}pectra {D}atabase, doi:10.18434/T4W30F

\bibitem[{Lemmens {$et~al$.}(2019)Lemmens, Rap, Thunnissen, Mackie, Candian,
  Tielens, Rijs, \& Buma}]{Lemmens19}
Lemmens, A.~K., Rap, D.~B., Thunnissen, J. M.~M., {$et~al$.} 2019, {A}\&{A},
  628, A130

\bibitem[{Malloci {$et~al$.}(2007)Malloci, Joblin, \& Mulas}]{Malloci07}
Malloci, G., Joblin, C., \& Mulas, G. 2007, {C}hem. {P}hys., 332, 353

\bibitem[{Malloci {$et~al$.}(2004)Malloci, Mulas, \& Joblin}]{Malloci04}
Malloci, G., Mulas, G., \& Joblin, C. 2004, {A}\&{A}, 426, 105

\bibitem[{Manfroid {$et~al$.}(2021)Manfroid, Hutsem{\'e}kers, \&
  Jehin}]{Manfroid21}
Manfroid, J., Hutsem{\'e}kers, D., \& Jehin, E. 2021, {N}ature, 593, 372

\bibitem[{Marsden(1994)}]{IAUC6093}
Marsden, B.~G. 1994, {C}omet {M}achholz (1994r), {I}nternational {A}stronomical
  {U}nion {C}ircular {N}o. 6093

\bibitem[{{Mazzotta Epifani} {$et~al$.}(2009){Mazzotta Epifani}, Palumbo, \&
  Colangeli}]{MazzottaEpifani09}
{Mazzotta Epifani}, E., Palumbo, P., \& Colangeli, L. 2009, {A}\&{A}, 508, 1031

\bibitem[{{McCandliss} {$et~al$.}(2010){McCandliss}, France, Osterman, Green,
  {McPhate}, \& Wilkinson}]{McCandliss10}
{McCandliss}, S.~R., France, K., Osterman, S., {$et~al$.} 2010, {ApJL}, 709,
  L183

\bibitem[{{Medallon, S., Rickman, E., Brown, J., et al.}(2023)}]{Medallon23}
{Medallon, S., Rickman, E., Brown, J., et al.} 2023, {STIS} {I}nstrument
  {H}andbook, {V}ersion 23.0, {B}altimore: {STScI}

\bibitem[{Moreels {$et~al$.}(1994)Moreels, Clairemidi, Hermine, Br{\'e}chignac,
  \& Rousselot}]{Moreels94}
Moreels, G., Clairemidi, J., Hermine, P., Br{\'e}chignac, P., \& Rousselot, P.
  1994, {A}\&{A}, 282, 643

\bibitem[{Mount \& Rottman(1981)}]{Mount81}
Mount, G.~H., \& Rottman, G.~J. 1981, {J}. {G}eophys. {R}es., 86, 9193

\bibitem[{Mulliken(1955)}]{Mulliken55}
Mulliken, R.~S. 1955, {J}. {C}hem. {P}hys., 23, 1997

\bibitem[{Pariser(1956)}]{Pariser56}
Pariser, R. 1956, {J}. {C}hem. {P}hys., 24, 250

\bibitem[{Peck \& Reeder(1972)}]{Peck72}
Peck, E.~R., \& Reeder, K. 1972, {J}. {O}pt. {S}oc. {A}m., 62, 958

\bibitem[{Rubin {$et~al$.}(2022)Rubin, Altwegg, Berthelier, Combi, {De Keyser},
  Dhooghe, Fuselier, Gombosi, H{\"a}nni, M{\"u}ller, Pestoni, Wampfler, \&
  Wurz}]{Rubin22}
Rubin, M., Altwegg, K., Berthelier, J.-J., {$et~al$.} 2022, {A}\&{A}, 658, A87

\bibitem[{Sandford {$et~al$.}(2006)Sandford, Al{\'e}on, Alexander, Araki, Bajt,
  Baratta, Borg, Bradley, Brownlee, Brucato, Burchell, Busemann, Butterworth,
  Clemett, Cody, Colangeli, Cooper, {D'Hendecourt}, Djouadi, Dworkin, Ferrini,
  Fleckenstein, Flynn, Franchi, Fries, Gilles, Glavina, Gounelle, Grossemy,
  Jacobsen, Keller, Kilcoyne, Leitner, Matrajt, Meibom, Mennella, Mostefaoui,
  Nittler, Palumbo, Papanastassiou, Robert, Rotundi, Snead, Spencer,
  Stadermann, Steele, Stephan, Tsou, Tyliszczak, Westphal, Wirick, Wopenka,
  Yabuta, Zare, \& Zolensky}]{Sandford06}
Sandford, K.~A., Al{\'e}on, J., Alexander, C. M.~O., {$et~al$.} 2006,
  {Science}, 314, 1720

\bibitem[{Schuhmann {$et~al$.}(2019)Schuhmann, Altwegg, Balsiger, Berthelier,
  {De Keyser}, Fiethe, Fuselier, Gasc, Gombosi, H{\"a}nni, Rubin, Tzou, \&
  Wampfler}]{Schuhmann19}
Schuhmann, M., Altwegg, K., Balsiger, H., {$et~al$.} 2019, {A}\&{A}, 630, A31

\bibitem[{Selco \& Carrick(1989)}]{Selco89}
Selco, J.~I., \& Carrick, P.~G. 1989, {J}. {M}ol. {S}pectrosc., 137, 13

\bibitem[{Sunuwar \& Manzanares(2021)}]{Sunuwar21}
Sunuwar, S., \& Manzanares, C.~E. 2021, {I}carus, 370, 114689

\bibitem[{Szczepanski {$et~al$.}(1995)Szczepanski, Wehlburg, \&
  Vala}]{Szczepanski95}
Szczepanski, J., Wehlburg, C., \& Vala, M. 1995, {C}hem. {P}hys. {L}ett., 232,
  221

\bibitem[{Tozzi {$et~al$.}(1998)Tozzi, Feldman, \& Festou}]{Tozzi98}
Tozzi, G.~P., Feldman, P.~D., \& Festou, M.~C. 1998, {A}\&{A}, 330, 753

\bibitem[{Venkataraman {$et~al$.}(2023)Venkataraman, Roy, Ramachandran,
  {Quiti{\'a}n-Lara}, Hill, Rajasekhar, Bhardwaj, Mason, \&
  Sivaraman}]{Venkataraman23}
Venkataraman, V., Roy, A., Ramachandran, R., {$et~al$.} 2023, {J}. {A}strophys.
  {A}stron., 444, 89

\bibitem[{Weaver {$et~al$.}(1981)Weaver, Feldman, Festou, {A'Hearn}, \&
  Keller}]{Weaver81}
Weaver, H.~A., Feldman, P.~D., Festou, M.~C., {A'Hearn}, M.~F., \& Keller,
  H.~U. 1981, {I}carus, 47, 449

\bibitem[{Weaver {$et~al$.}(1997)Weaver, Feldman, {A'Hearn}, Arpigny, Brandt,
  Festou, Haken, {McPhate}, Stern, \& Tozzi}]{Weaver97}
Weaver, H.~A., Feldman, P.~D., {A'Hearn}, M.~F., {$et~al$.} 1997, {S}cience,
  275, 1900

\bibitem[{Woods {$et~al$.}(1996)Woods, Prinz, Rottman, London, Crane, Cebula,
  Hilsenrath, Brueckner, Andrews, White, {VanHoosier}, Floyd, Herring, Knapp,
  Pankratz, \& Reiser}]{Woods96}
Woods, T.~N., Prinz, D.~K., Rottman, G.~J., {$et~al$.} 1996, {J}. {G}eophys.
  {R}es., 101, 9541

\end{thebibliography}
\balance
\end{document}